\documentclass[a4paper]{jpconf}
\usepackage{graphicx}

\begin{document}
\title{Low-Mass Dielectron Production in pp, p--Pb and Pb--Pb Collisions with ALICE}

\author{Patrick Reichelt (for the ALICE Collaboration)}

\address{Institut f\"ur Kernphysik, Goethe-Universit\"at Frankfurt am Main, Germany}

\ead{preichelt@ikf.uni-frankfurt.de}

\begin{abstract}
The ALICE Collaboration measures the production of low-mass dielectrons in pp, p--Pb and Pb--Pb collisions at the LHC. The main detectors used in the analyses are the Inner Tracking System, Time Projection Chamber and Time-Of-Flight detector, all located around mid-rapidity. The production of virtual photons relative to the inclusive yield in pp collisions is determined by analyzing the dielectron excess with respect to the expected hadronic sources. The direct photon cross section is then calculated and found to be in agreement with NLO pQCD calculations. Results from the invariant mass analysis in p--Pb collisions show an overall agreement between data and hadronic cocktail. In Pb--Pb collisions, uncorrected background-subtracted yields have been extracted in two centrality classes. A feasibility study for LHC run 3 after the ALICE upgrade indicates the possibility for a future measurement of the early effective temperature.
\end{abstract}

\section{Introduction}
The measurement of electron-positron pairs (dielectrons) in the low invariant mass region allows studying the vacuum and in-medium properties of light vector mesons. Additionally, low-mass dielectrons are produced by internal conversion of virtual direct photons. They are excellent probes to study all collision stages, since they pass through the created medium almost unaffected. To quantify modifications of the dielectron production in heavy-ion collisions, measurements in pp collisions serve as a reference, while the analysis of p-A collisions allows disentangling cold from hot nuclear matter effects. In ALICE \cite{Ali-Exp} at the LHC, dielectron measurements are performed using the central barrel detectors around mid-rapidity. Electrons can be identified via their specific energy loss in the Inner Tracking System (ITS) and the Time Projection Chamber (TPC), combined with time-of-flight information from the TOF detector \cite{Ali-Perf}. In this proceedings we present a virtual direct photon measurement in pp collisions and the invariant mass analyses in p--Pb and Pb--Pb collisions. Prospects of a future measurement in Pb--Pb after the ALICE upgrade for LHC run 3 are also discussed.

\section{Virtual direct photon production in pp collisions at $ \sqrt{s} = 7 {\rm ~TeV}$ }
\label{sect-pp}
\begin{figure}[h]
\begin{minipage}{20.6pc}
\vspace{0.2pc}
\includegraphics[width=20pc]{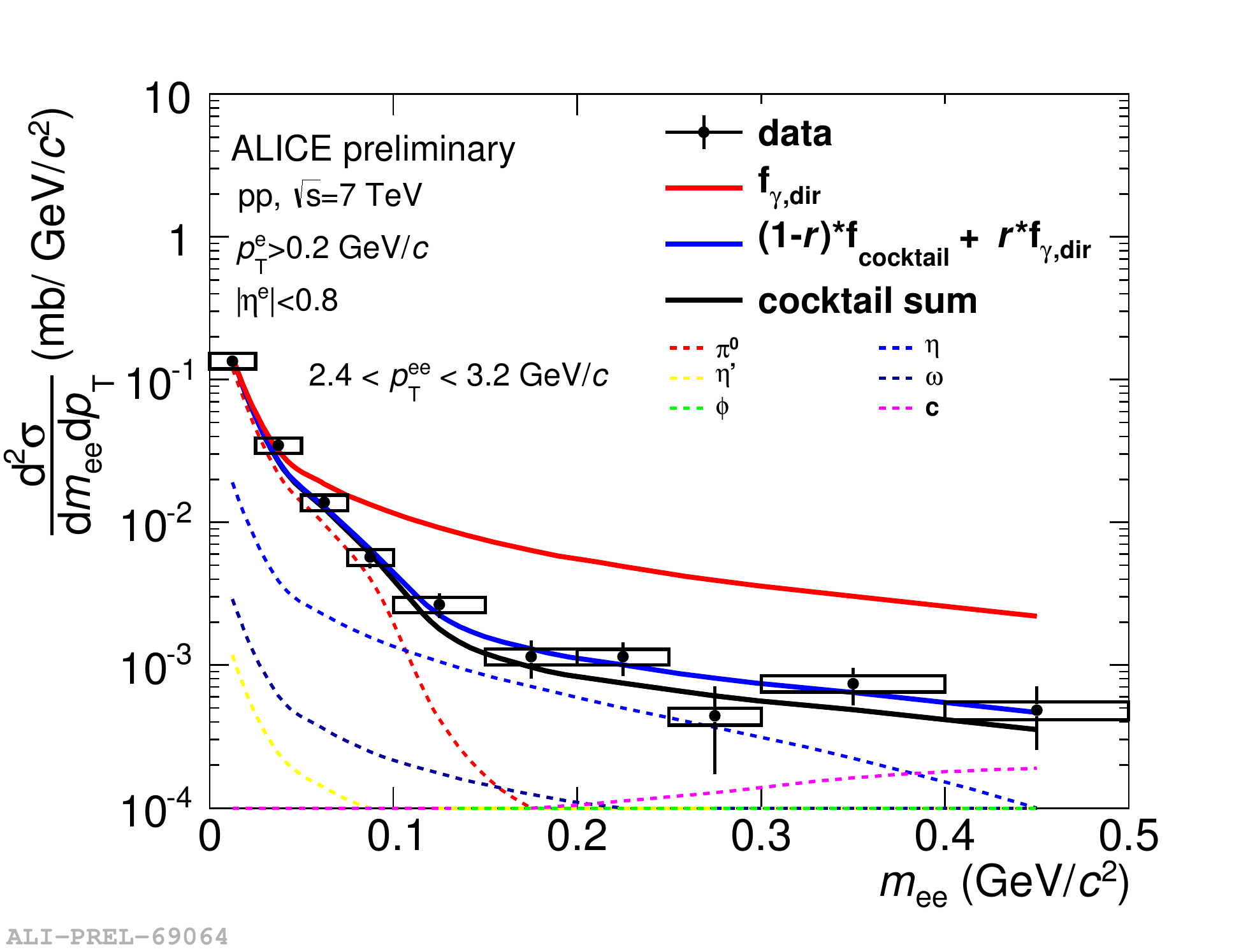}
\caption{\label{pp-fit}Low-mass region of the dielectron cross section for $2.4 < p_{\rm T}^{\rm ee} < 3.2 {\rm ~GeV}/c$ compared to the hadronic cocktail, as well as to a fit to extract the virtual direct photon fraction $r$.}
\end{minipage}\hspace{2pc}%
\begin{minipage}{14pc}
\includegraphics[width=13.8pc]{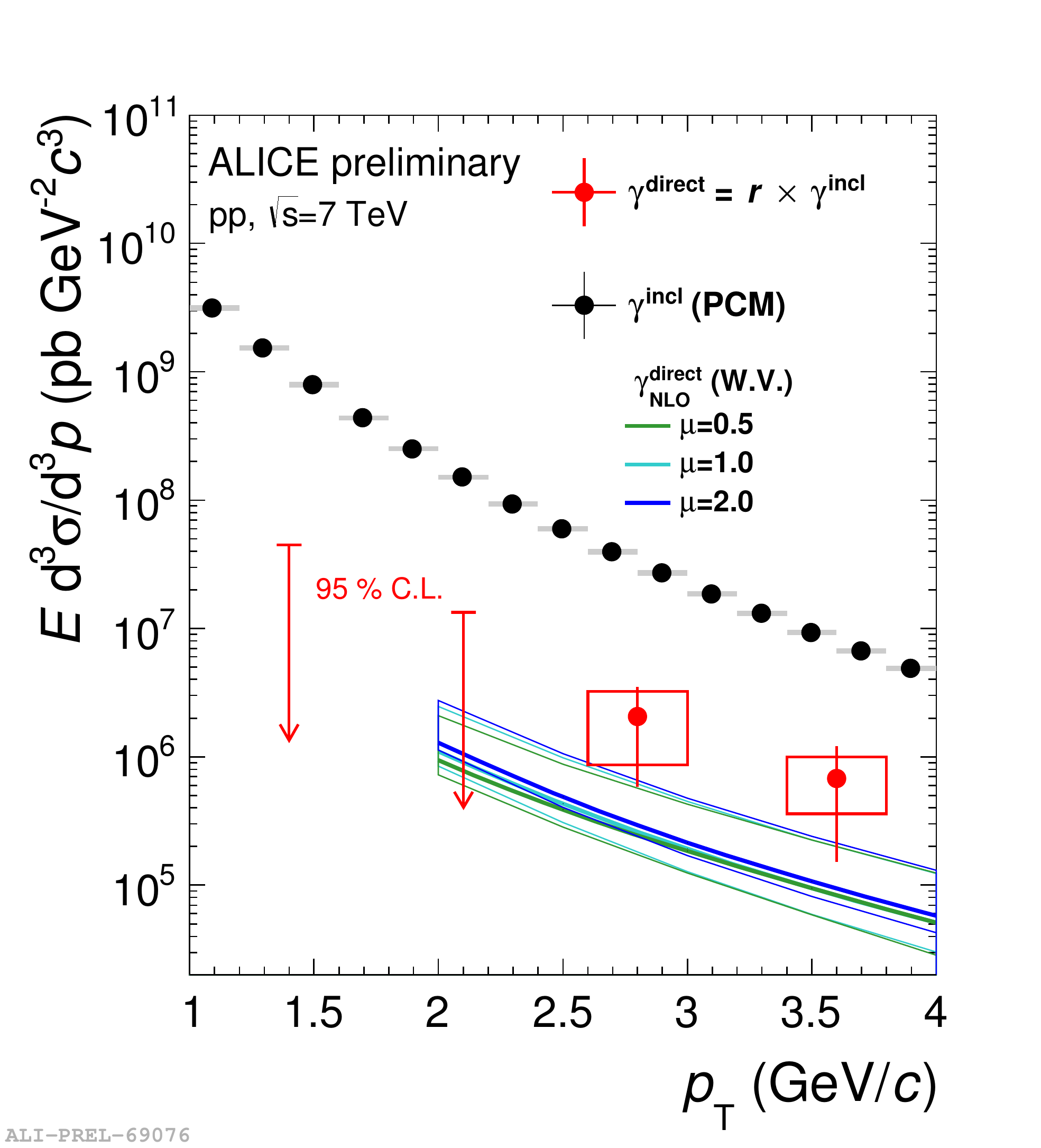}
\caption{\label{pp-photons}Direct photon cross section compared to NLO pQCD calculations, and the inclusive photon cross section extracted from PCM.}
\end{minipage} 
\end{figure}

The analysis presented here is based on $3 \cdot 10^{8}$ minimum bias pp collisions at $ \sqrt{s} = 7 {\rm ~TeV}$, recorded in 2010. Fiducial cuts on transverse momentum ($p_{\rm T} > 0.2 {\rm ~GeV}/c$) and pseudorapidity ($ |\eta| < 0.8$) are applied to electron candidates. A clean electron sample is achieved by cuts on the time-of-flight and on the ${\rm d}E/{\rm d}x$ in the TPC. Dielectron spectra are created for unlike-sign and like-sign combinations of these particles. The unlike-sign distribution contains a superposition of signal and combinatorial background, while the like-sign is used to provide an estimate of such a background. The acceptance may be different for unlike- and like-sign pairs due to detector effects. Therefore, the signal yield is obtained as $N_{\rm ULS}^{\rm same} - N_{\rm LS}^{\rm same} \cdot R$, where $R$ is built via a mixed-event technique as $R = N_{\rm ULS}^{\rm mix} / N_{\rm LS}^{\rm mix}$.\\
Figure \ref{pp-fit} shows the differential dielectron cross-section in a kinematic region $p_{\rm T}^{\rm ee} \gg m_{\rm ee}$ which is useful to study direct photon production. Also shown in Fig. \ref{pp-fit} are a hadronic cocktail and its components as well as the expected mass distribution of dielectrons coming from virtual direct photons, after having applied the single-electron fiducial cuts on $p_{\rm T}$ and $\eta$. A fit to the data is performed to determine the virtual photon fraction $r$ by using the function $f_{\rm combined} = (1-r) \cdot f_{\rm cocktail} + r \cdot f_{\rm photon}$ in the range $0.1 < m_{\rm ee} < 0.4 {\rm ~GeV}/c^2$. This fit is done in four pair-$p_{\rm T}$ regions and the extracted virtual photon fraction is then multiplied by the inclusive photon cross section, also measured by ALICE via the photon conversion method (PCM) \cite{Ali-PCM}. In Figure \ref{pp-photons} the resulting direct photon cross section is compared to NLO pQCD calculations. Reasonable agreement is found, within uncertainties, between data and model. Details on this analysis can be found in \cite{Koehler}.

\section{Low-mass dielectron production in p--Pb collisions at $ \sqrt{s_{\rm NN}} = 5.02 {\rm ~TeV}$ }
\begin{figure}[h]
\begin{minipage}{18pc}
\vspace{0.3pc}
\includegraphics[width=17pc]{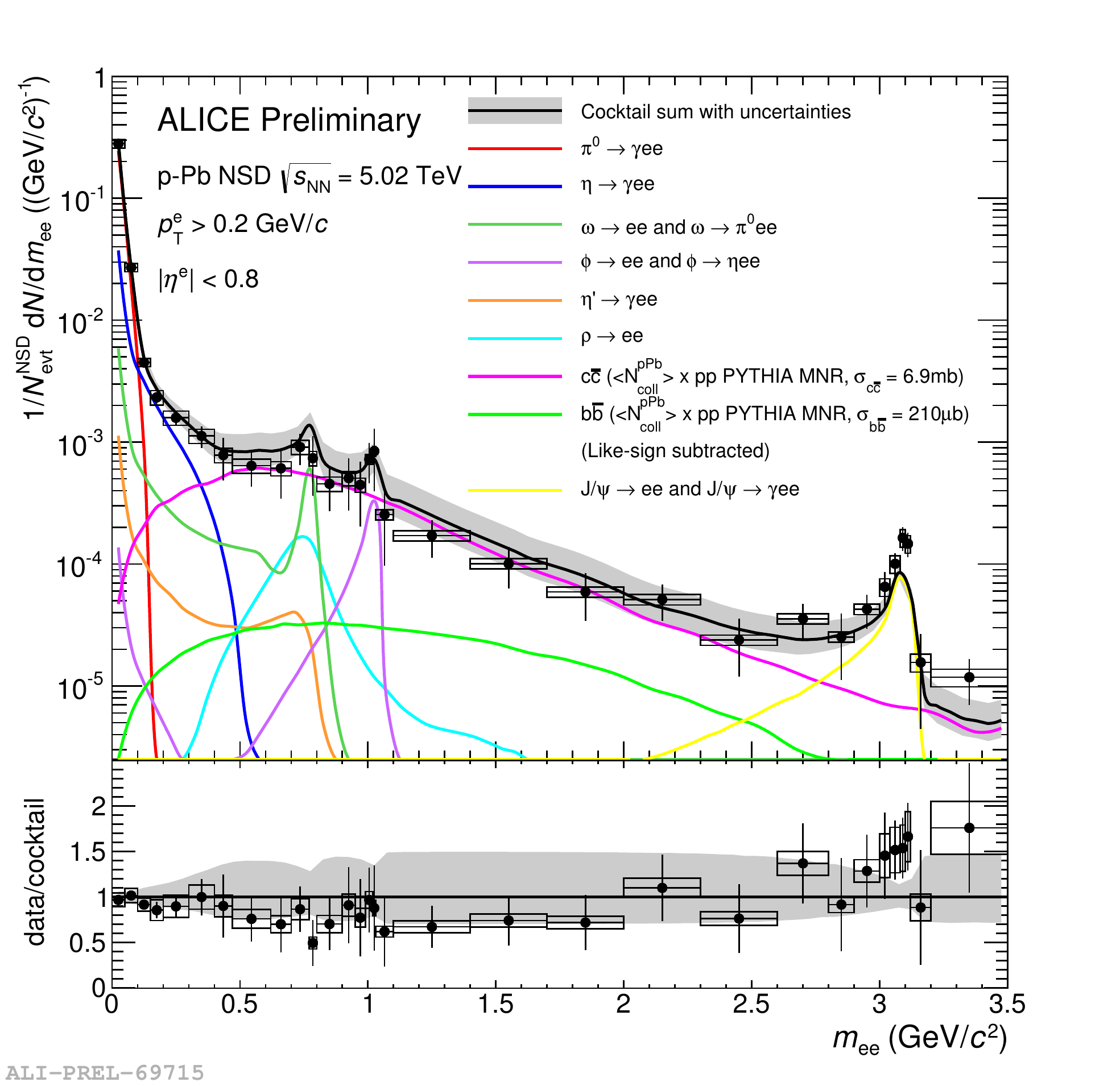}
\caption{\label{p-Pb-mee}Dielectron mass spectrum from p--Pb collisions in comparison to the hadronic cocktail.}
\end{minipage}\hspace{2pc}%
\begin{minipage}{18pc}
\includegraphics[width=17pc]{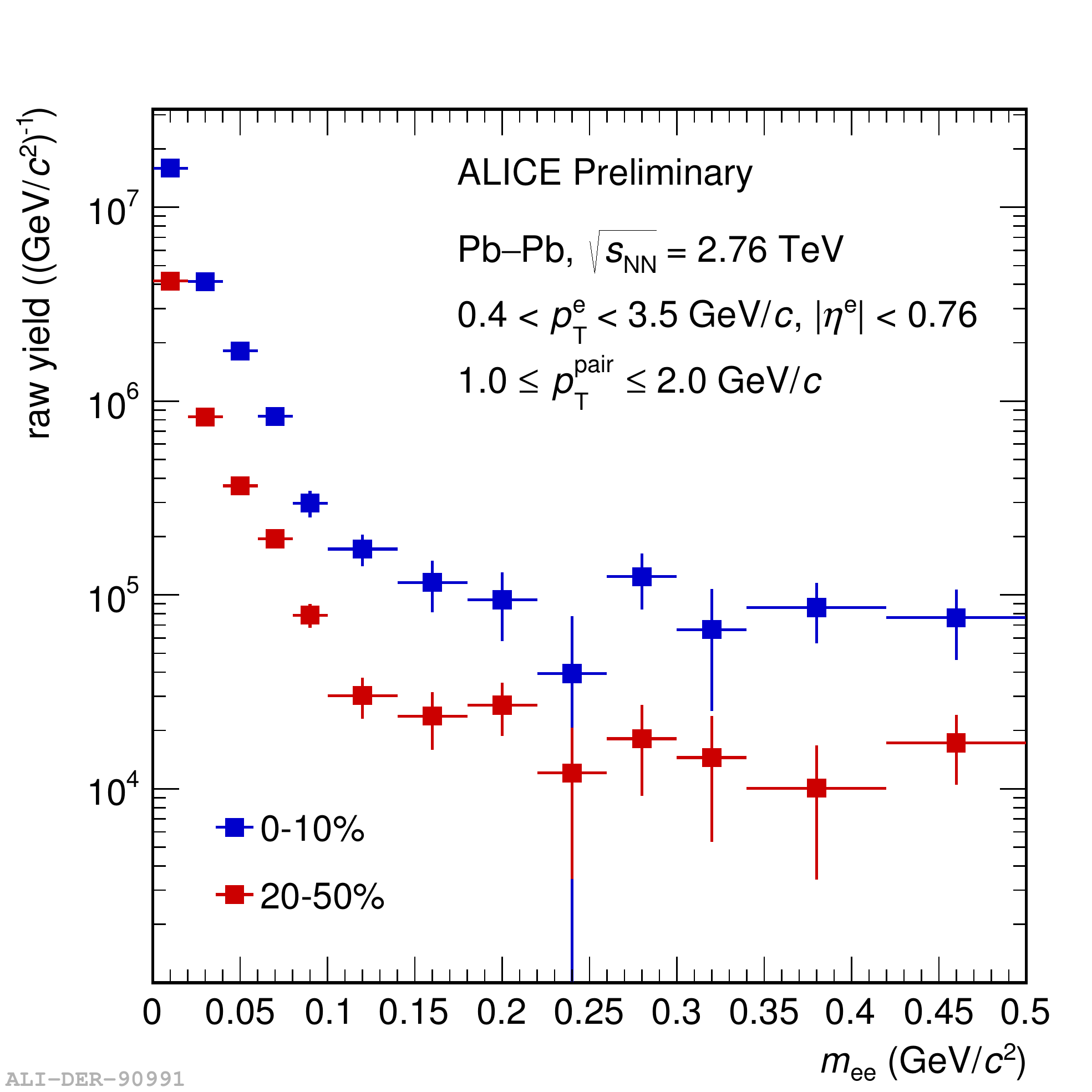}
\caption{\label{Pb-Pb-mee}Dielectron raw subtracted yield in central and semi-central Pb--Pb collisions after background subtraction.}
\end{minipage} 
\end{figure}

The analyzed data sample consists of $10^{8}$ minimum bias p--Pb collisions. The analysis is carried out in $ |\eta| < 0.8$ and two different track selections $p_{\rm T} > 0.2 {\rm ~GeV}/c$ or $p_{\rm T} > 0.4 {\rm ~GeV}/c$ are performed, to enable comparisons with both the pp and Pb--Pb data. The PID scheme was modified to increase the electron efficiency: a TOF signal is not required, yet used if available to purify the electron sample. Kaons and protons are instead rejected by an electron inclusion cut on the ${\rm d}E/{\rm d}x$ measured by the Silicon Drift and Silicon Pixel Detectors of the ITS. In addition, electrons from photon conversions are identified and rejected either by their displaced vertex or by the orientation of the pair plane relative to the magnetic field. More in detail, since a conversion pair does not have an intrinsic opening angle, an apparent opening after reconstruction may only appear perpendicular to the magnetic field vector, while true finite-mass signal pairs have a uniform azimuthal opening angle. Therefore, an exclusion cut on the pair plane normal vector $\phi_{\rm V} > 2.5$ is applied for $m_{\rm ee} < 0.05 {\rm ~GeV}/c^2$.\\
The signal extraction is performed as described in section \ref{sect-pp}. The signal-to-background ratio is smaller than in pp data by a factor two to four, and range from $50$ in the $\pi^0$ mass region to $0.02$ at $m_{\rm ee} \approx 0.5 {\rm ~GeV}/c^2$ and to $\approx 4$ in the $J/\psi$ region. The corrected dielectron yield for $p_{\rm T} > 0.2 {\rm ~GeV}/c$ and integrated over pair-$p_{\rm T}$, is shown in Fig. \ref{p-Pb-mee}. The data points and systematic uncertainties are extracted from the mean values and the spread of results obtained with $22$ different combinations of analysis cut settings (inspired by \cite{Barlow}). The data are compared to the hadronic cocktail, which uses the charged pion measurement \cite{Ali-pi-charged} as $\pi^0$ input and $m_{\rm T}$-scaling for the other light mesons. Open heavy-flavour and $J/\psi$ contributions are calculated from PYTHIA, tuned to independent ALICE measurements in pp and p--Pb. Uncertainties on the input of the cocktail sources are propagated to the shapes shown in Fig. \ref{p-Pb-mee}. Data and cocktail are in reasonable agreement within their uncertainties over the full mass range.

\section{Low-mass dielectron continuum in Pb--Pb collisions at $ \sqrt{s_{\rm NN}} = 2.76 {\rm ~TeV}$ }
From the 2011 Pb--Pb run, $1.7 \cdot 10^{7}$ and $1.2 \cdot 10^{7}$ events in central (0-10\%) and semi-central (20-50\%) Pb--Pb collisions, respectively, are selected for this analysis. The PID scheme is as described for the p--Pb analysis. A track selection cut $0.4 < p_{\rm T} < 3.5 {\rm ~GeV}/c$ ensures exclusion of charged pions and reduces the contribution from soft conversion and $\pi^0$-Dalitz electrons to the combinatorial background. A residual hadron contamination at the $\approx 10\%$ level also contributes to the background pairs but Monte-Carlo studies show that it does not distort the signal. A conversion rejection cut $\phi_{\rm V} > 2.36$ is applied for $m_{\rm ee} < 0.1 {\rm ~GeV}/c^2$. The signal-to-background ratio for $0.2 <  m_{\rm ee} < 0.5 {\rm ~GeV}/c^2$ decreases from $0.01$ for 20-50\% to $0.003$ for 0-10\% centrality. The combinatorial-background-subtracted uncorrected yields (\textsl{raw subtracted yield}) are shown in Fig. \ref{Pb-Pb-mee} in the low-mass region for $1 < p_{\rm T}^{\rm ee} < 2 {\rm ~GeV}/c$ for both centralities. Efficiency correction and cocktail comparison are in progress.

\section{Upgrade Study for future dielectron measurements in Pb--Pb collisions}
\begin{figure}[h]
\begin{minipage}{18pc}
\vspace{0.2pc}
\includegraphics[width=16.7pc]{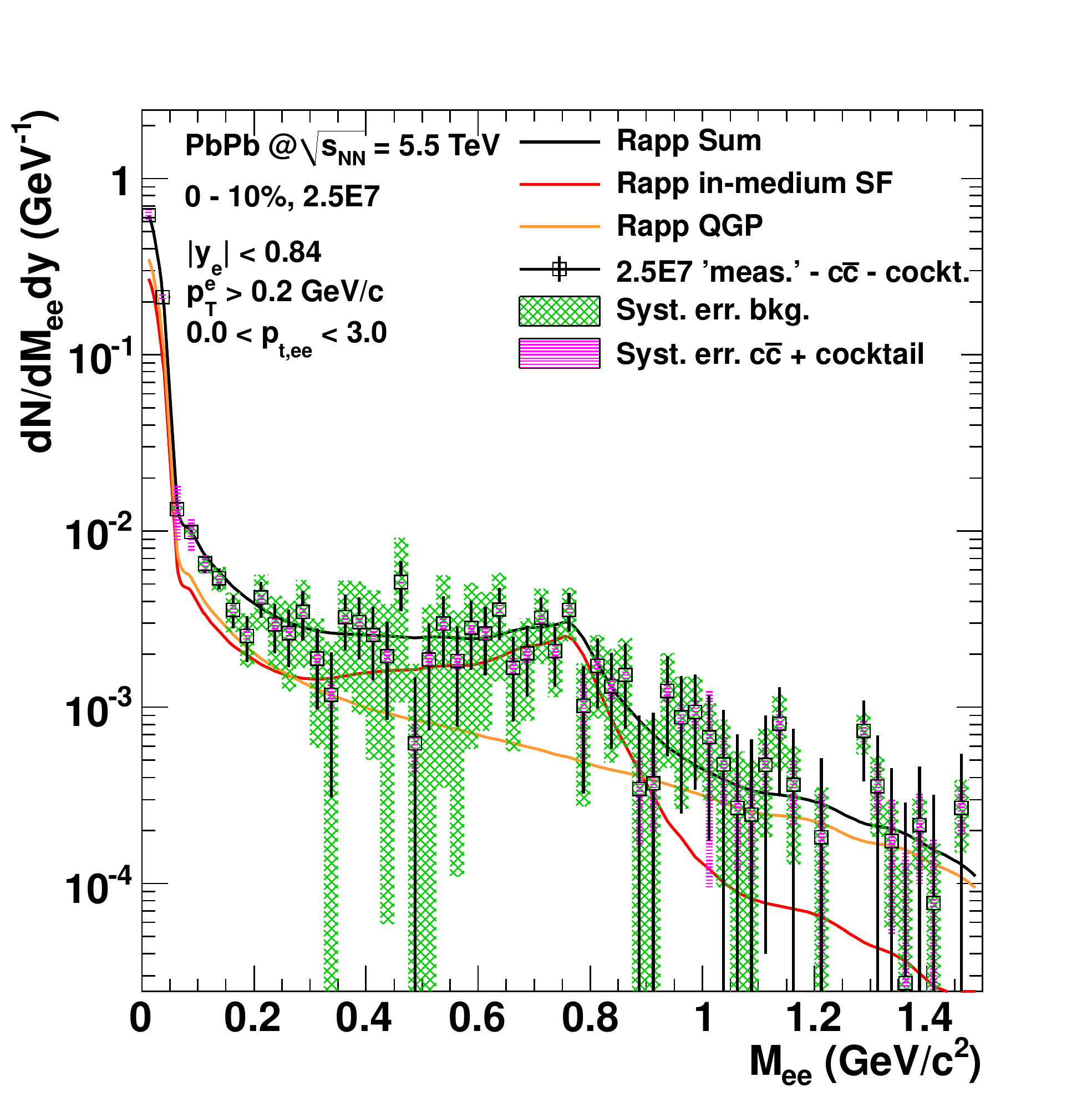}
\caption{\label{excess-curr}Expected dielectron excess yield for $ \sqrt{s_{\rm NN}} = 5.5 {\rm ~TeV}$ after hadronic cocktail subtraction, using the current experimental setup of ALICE.}
\end{minipage}\hspace{2pc}%
\begin{minipage}{18pc}
\includegraphics[width=16.7pc]{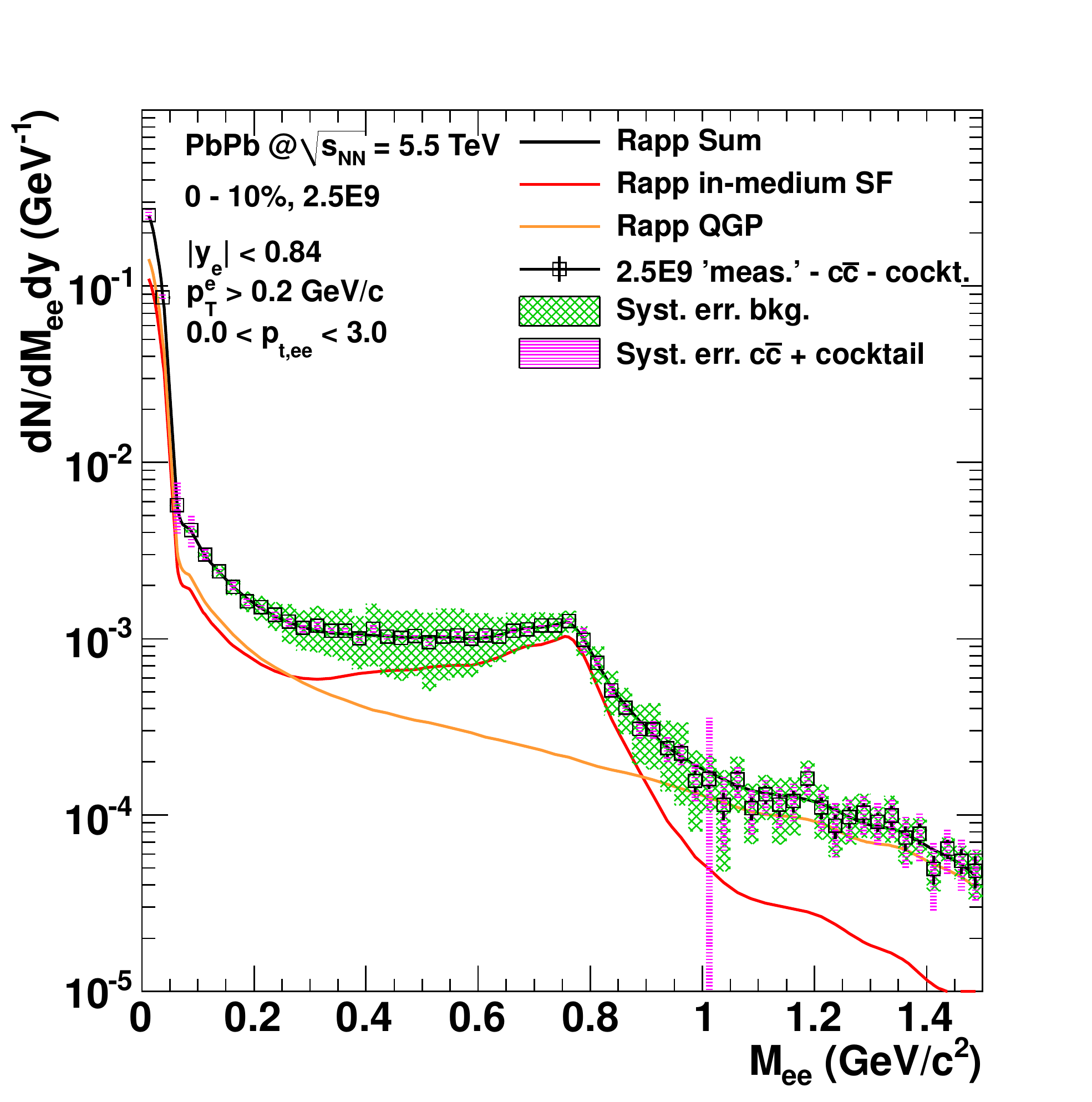}
\caption{\label{excess-upg}Same as Fig. \ref{excess-curr}, but with up\-gra\-ded ITS and TPC. Reduced statistical and systematic uncertainties make a temperature measurement in the IMR feasible.}
\end{minipage} 
\end{figure}

A fine-binning differential low-mass dielectron measurement in Pb--Pb collisions is a major physics case for the ALICE upgrade for LHC run 3 \cite{Ali-Upg-LOI}. We evaluated to which precision a possible thermal excess yield \cite{Rapp-1999} could be reconstructed in central and peripheral Pb--Pb collisions at full LHC energy ($ \sqrt{s_{\rm NN}}=$ 5.5 TeV). Electron tracking and PID efficiencies from a Monte-Carlo simulation using the current ALICE setup, operated at reduced solenoid magnetic field, are used and modified according to the expected performance after the upgrade. 
A statistical significance is calculated from realistically constructed signal and background spectra, and is then used for Poisson sampling of data points around the model input. Figures \ref{excess-curr} and \ref{excess-upg} show the expected spectra after subtraction of hadronic sources, for $2.5 \cdot 10^{7}$ and $2.5 \cdot 10^{9}$ events using the current and upgraded setup, respectively. While the measurement is dominated by un\-cer\-tain\-ties in the first case, the upgrade is expected to allow for a temperature measurement in the intermediate mass region (IMR) within $10\%$ statistical and $20\%$ systematic uncertainty \cite{Ali-Upg-ITS-TDR}.

\section{Conclusions}
Virtual direct photons are observed in pp collisions at $p_{\rm T}^{\rm ee} \gg m_{\rm ee}$. The extracted direct photon cross section agrees with NLO pQCD calculations within uncertainties. The pair-$p_{\rm T}$-integrated invariant mass spectrum measured in p--Pb collisions is in agreement with the expected hadronic sources. From Pb--Pb collisions, raw subtracted yields have been extracted in two centrality classes and a further analysis is ongoing. A feasibility study indicates that the ALICE upgrade for LHC run 3 will facilitate a measurement of the early effective temperature.

\end{document}